%% file: SMP-17-013_temp.tex
\pdfoutput=1

\documentclass[11pt,twoside,a4paper,cmspaper,final,collab]{cms-tdr}
\def\svnVersion{830fd98-D}\def\svnDate{2019/07/26}\def\cmsCernNoTag{CERN-EP-2019-074}\def\cmsCernDate{\today}\def\cmsMessage{"Published in Physical Review D as \href{http://dx.doi.org/10.1103/PhysRevD.100.012004}{\doi{10.1103/PhysRevD.100.012004}.}"}

\begin{document}\cmsNoteHeader{SMP-17-013}

\hyphenation{had-ron-i-za-tion}
\hyphenation{cal-or-i-me-ter}
\hyphenation{de-vices}
\ifthenelse{\boolean{cms@external}}{\providecommand{\cmsTable}[1]{#1}}{\providecommand{\cmsTable}[1]{\resizebox{\textwidth}{!}{#1}}}
 \newlength\cmsTabSkip\setlength{\cmsTabSkip}{1.5ex}
\ifthenelse{\boolean{cms@external}}{\providecommand{\NA}{\ensuremath{\cdots}\xspace}}{\providecommand{\NA}{\ensuremath{\text{---}}\xspace}}
\newcommand{\MTmax}{\ensuremath{\mT^{\text{max}}}\xspace}
\newcommand{\MTthird}{\ensuremath{\mT^{\text{3rd}}}\xspace}
\newcommand{\SSee}{\ensuremath{\Pepm\Pepm}\xspace}
\newcommand{\SSem}{\ensuremath{\Pepm\PGmpm}\xspace}
\newcommand{\SSmm}{\ensuremath{\PGmpm\PGmpm}\xspace}
\newcommand{\Mll}{\ensuremath{m_{\ell\ell}}\xspace}
\newcommand{\MSFOS}{\ensuremath{m_{\mathrm{SFOS}}}\xspace}
\newcommand{\ptlll}{\ensuremath{\pt(\ell\ell\ell)}\xspace}
\newcommand{\ptlllvec}{\ensuremath{\ptvec(\ell\ell\ell)}\xspace}
\newcommand{\DPhilllMET}{\ensuremath{\Delta\phi\left(\ptlllvec,\ptvecmiss\right)}\xspace}
\newcommand{\Mlll}{\ensuremath{m_{\ell\ell\ell}}\xspace}
\newcommand{\Mjj}{\ensuremath{m_{\mathrm{jj}}}\xspace}
\newcommand{\MjjL}{\ensuremath{m_{\mathrm{JJ}}}\xspace}
\newcommand{\DetaJJ}{\ensuremath{\Delta\eta_{\mathrm{JJ}}}\xspace}
\newcommand{\theLumi}{\ensuremath{35.9\fbinv}\xspace}
\newcommand{\Wmp}{\PWmp}
\newcommand{\WH}{\ensuremath{\PW\PH}\xspace}
\newcommand{\WZ}{\ensuremath{\PW\PZ}\xspace}
\newcommand{\ZZ}{\ensuremath{\PZ\PZ}\xspace}
\newcommand{\WWWpm}{\ensuremath{\Wpm\Wpm\Wmp}\xspace}
\newcommand{\WWW}{\ensuremath{\PW\PW\PW}\xspace}
\newcommand{\WHtoWWW}{\ensuremath{\WH\to\WWW^{*}}\xspace}
\newcommand{\WW}{\ensuremath{\PW\PW}\xspace}
\newcommand{\SSWW}{\ensuremath{\Wpm\Wpm}\xspace}
\newcommand{\ttW}{\ensuremath{\ttbar\Wpm}\xspace}
\newcommand{\ttZ}{\ensuremath{\ttbar\PZ}\xspace}
\newcommand{\ttV}{\ensuremath{\ttbar\mathrm{V}}\xspace}
\newcommand{\Wg}{\ensuremath{\PW\gamma}\xspace}
\newcommand{\eTL}{\ensuremath{\epsilon_{\mathrm{TL}}}\xspace}
\newcommand{\ptl}{\ensuremath{\pt^{\ell}}\xspace}
\newcommand{\ptlvec}{\ensuremath{\ptvec^{\,\ell}}\xspace}
\newcommand{\ptjetvec}{\ensuremath{\ptvec^{\,\text{jet}}}\xspace}
\newcommand{\ptcor}{\ensuremath{\pt^{\text{corr}}}\xspace}
\newcommand{\MZ}{\ensuremath{m_{\PZ}}\xspace}
\newcommand{\IP} {\ensuremath{b}\xspace}
\newcommand{\IREL} {\ensuremath{I_{\text{rel}}}\xspace}
\newcommand{\hatsWWW}{\ensuremath{\hat{s}_{\PW\PW\PW}}\xspace}
\newcommand{\ST}{\ensuremath{S_{\mathrm{T}}}\xspace}
\newcommand{\STmin}{\ensuremath{S_{\mathrm{T}}^{\text{min}}}\xspace}
\newcommand{\Aeff}{\ensuremath{A_{\text{eff}}}\xspace}
\newcommand{\fTa}{\ensuremath{f_{\mathrm{T},0}}\xspace}
\newcommand{\fTb}{\ensuremath{f_{\mathrm{T},1}}\xspace}
\newcommand{\fTc}{\ensuremath{f_{\mathrm{T},2}}\xspace}
\newcommand{\ptPU}{\ensuremath{\pt^{\mathrm{PU}}}\xspace}
\newcommand{\ptnc}{\ensuremath{\pt^{\mathrm{nc}}}\xspace}
\newcommand{\ptc}{\ensuremath{\pt^{\mathrm{c}}}\xspace}
\newcommand{\Palp}{\ensuremath{\mathrm{a}}\xspace}
\newcommand{\malp}{\ensuremath{m_{\Palp}}\xspace}
\newcommand{\falp}{\ensuremath{f_{\Palp}}\xspace}
\newcommand{\mW}{\ensuremath{m_{\PW}}\xspace}
\newcommand{\muR}{\ensuremath{\mu_{\mathrm{R}}}\xspace}
\newcommand{\muF}{\ensuremath{\mu_{\mathrm{F}}}\xspace}
\newcommand{\LTERM}[1] { \frac{f_{\mathrm{#1}}}{\Lambda^4} {\mathcal{O}}_{\mathrm{#1}}}
\ifthenelse{\boolean{cms@external}}{\providecommand{\CL}{C.L.\xspace}}{\providecommand{\CL}{CL\xspace}}
\ifthenelse{\boolean{cms@external}}{\providecommand{\NA}{\ensuremath{\cdots}\xspace}}{\providecommand{\NA}{\ensuremath{\text{---}}\xspace}}

\RCS$Revision$
\RCS$HeadURL$
\RCS$Id$
\newlength\cmsFigWidth
\ifthenelse{\boolean{cms@external}}{\setlength\cmsFigWidth{0.85\columnwidth}}{\setlength\cmsFigWidth{0.4\textwidth}}
\ifthenelse{\boolean{cms@external}}{\providecommand{\cmsLeft}{upper\xspace}}{\providecommand{\cmsLeft}{left\xspace}}
\ifthenelse{\boolean{cms@external}}{\providecommand{\cmsRight}{lower\xspace}}{\providecommand{\cmsRight}{right\xspace}}
\cmsNoteHeader{SMP-17-013}
\title{Search for the production of \texorpdfstring{\WWWpm events at $\sqrt{s} = 13\TeV$}{WWW events at sqrt(s) = 13 TeV}}

\date{\today}

\abstract{
A search for the production of events containing three \PW bosons predicted by the standard model is reported. The search
is based on a data sample of proton-proton collisions at a center-of-mass energy of 13\TeV
recorded by the CMS experiment at the CERN LHC and corresponding to a total integrated luminosity of
\theLumi. The search is performed in final states with three leptons (electrons or muons),
or with two same-charge leptons plus two jets.
The observed (expected) significance of the signal for \WWWpm production is $0.60~(1.78)$ standard deviations,
and the ratio of the measured signal yield to that expected from the standard model
is $0.34^{+0.62}_{-0.34}$.
Limits are placed on three anomalous quartic gauge couplings and on the production of massive axionlike particles.
}
\hypersetup{%
pdfauthor={CMS Collaboration},%
pdftitle={Search for the production of WWW events at sqrt(s) = 13 TeV},%
pdfsubject={CMS},%
pdfkeywords={CMS, physics}}

\maketitle
\section{Introduction\label{sec:intro}}

\begin{figure*}[htb]
\centering
    \includegraphics[width=0.245\textwidth]{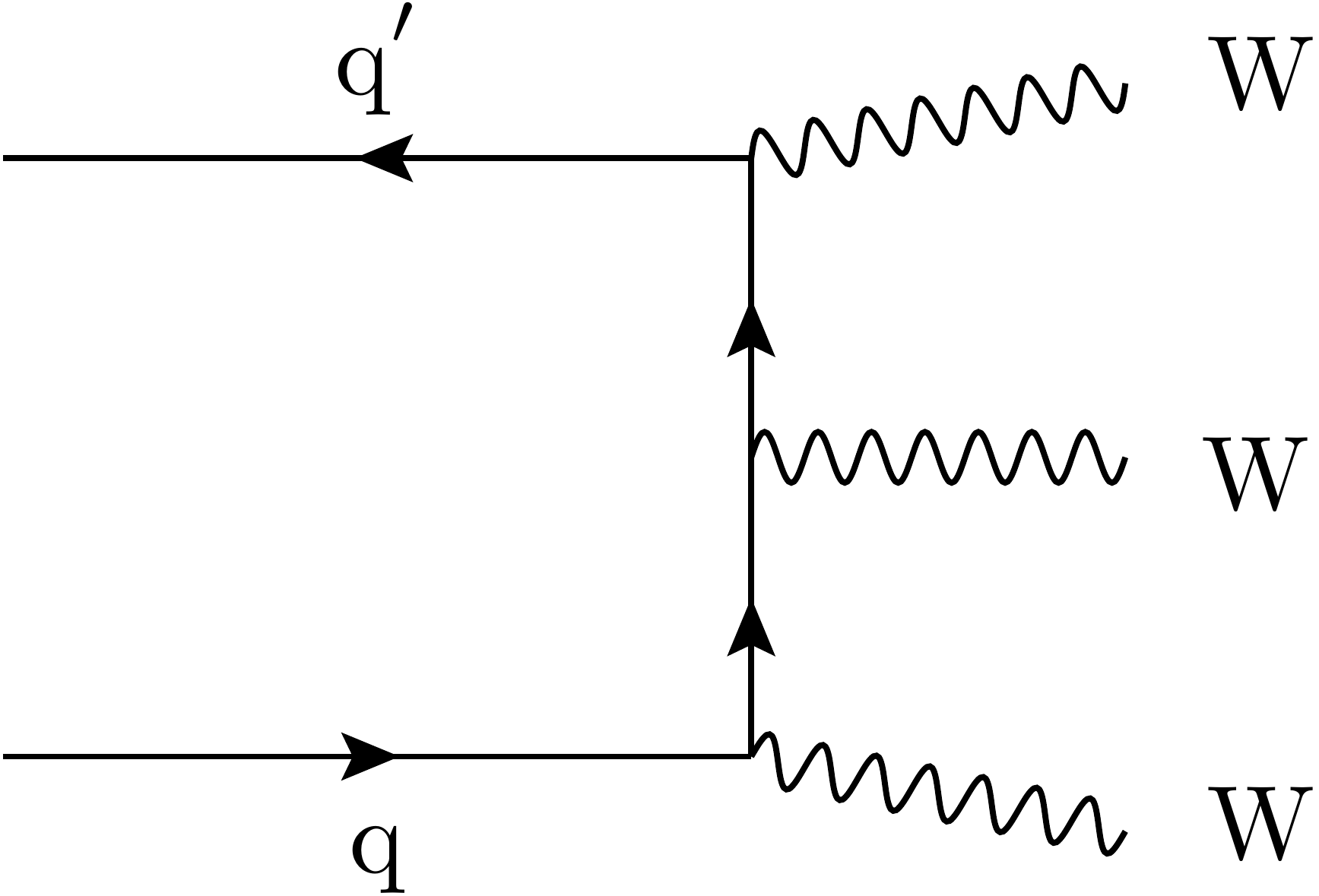}
    \includegraphics[width=0.245\textwidth]{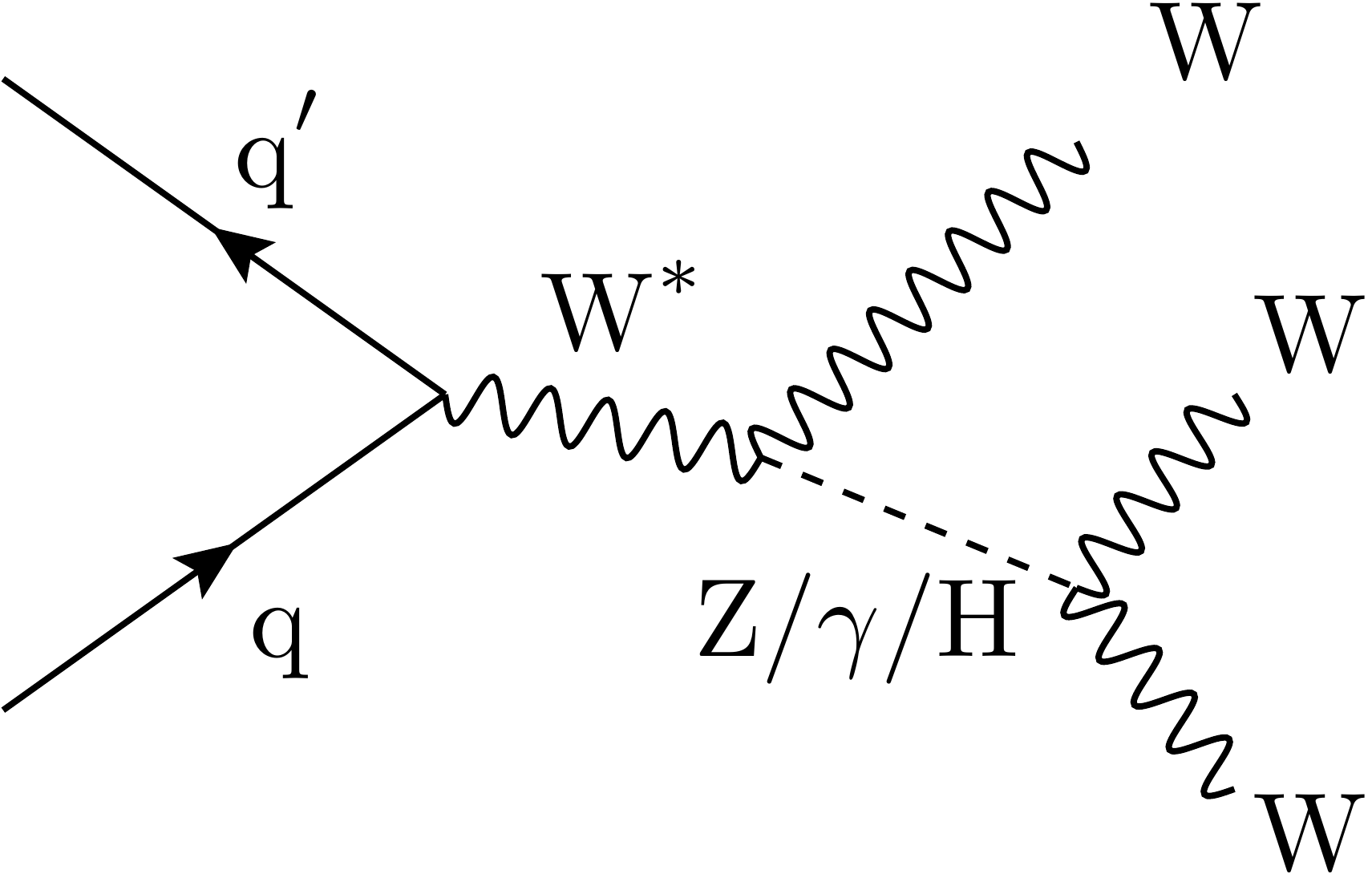}
    \includegraphics[width=0.245\textwidth]{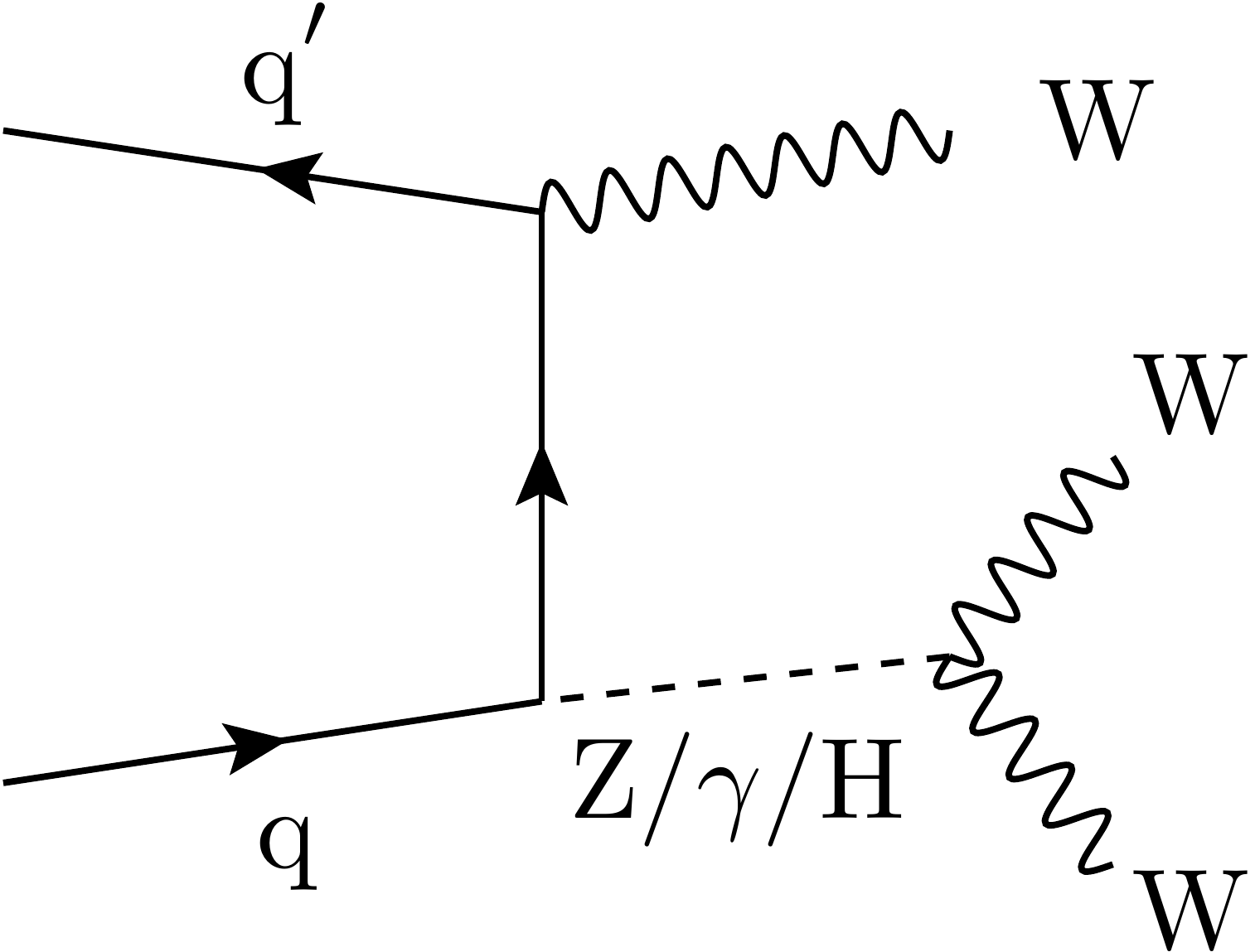}
    \includegraphics[width=0.245\textwidth]{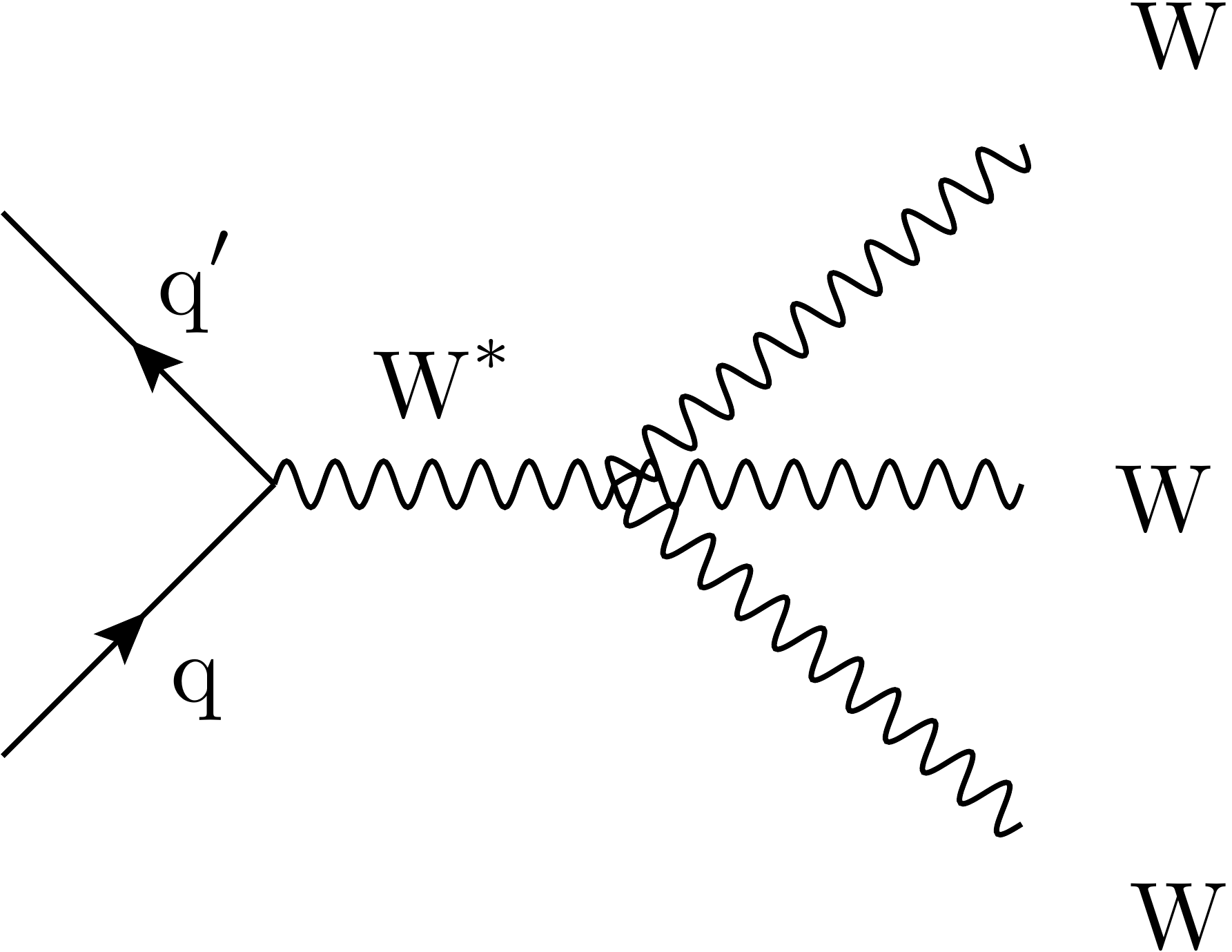}
    \caption{\label{fig:WWW:feynman}Tree-level Feynman diagrams for \WWW production}
\end{figure*}

According to the standard model (SM), events with three \PW bosons (\WWWpm, labeled \WWW in the following)
are produced in proton-proton ($\Pp\Pp$) collisions at the CERN LHC.
The process is sensitive to triple and quartic gauge couplings (QGC), so
the observation and study of this process provides an important test of the electroweak
sector of the SM.
Figure~\ref{fig:WWW:feynman} shows examples of lowest-order Feynman diagrams for \WWW production.
The analysis presented here focuses on the electroweak production of
\WWW events. The associated production of the Higgs (\PH) boson with a \PW boson,
where the \PH boson decays to $\PWp\PWm$, is considered to be part of the signal production, whereas other processes
such as the production of \ttW are considered to be background processes. The nonresonant \WWW production cross section
is calculated to be $216\pm9\unit{fb}$~\cite{Dittmaier:2017bnh} and,
after including the contribution of \WHtoWWW with one off-shell \PW boson~\cite{deFlorian:2016spz},
the total theoretical electroweak production cross section is $509\pm13\unit{fb}$.
In this paper, the label \WWW includes both types of production.

A search for \WWW production in 8\TeV $\Pp\Pp$ collision data~\cite{Aaboud:2016ftt} and
evidence for the production of three massive gauge bosons in 13\TeV $\Pp\Pp$ collisions~\cite{Aad:2019udh} were reported by the ATLAS Collaboration.

The analysis presented in this paper is performed with a sample of $\Pp\Pp$ collisions at a center-of-mass energy of 13\TeV
produced by the LHC and recorded with the CMS detector in 2016; the integrated luminosity
for this sample is \theLumi.

Events containing three \PW bosons can be classified by the expected number of charged leptons
(electrons or muons only) in the final state: 41.7\% contain no leptons, 42.4\% contain one lepton,
9.6\% have two leptons with opposite-sign (OS) charge, 4.8\% have two same-sign (SS)
leptons, and 1.6\% of all events contain three leptons ($3\ell$).
These branching fractions include the contributions from leptonic decays of $\tau$ leptons to electrons or muons and  neutrinos.
Large backgrounds from the production of events with multiple jets, \PW{}~bosons and jets, Drell-Yan
lepton pairs and jets, and \ttbar final states preclude the isolation of a signal except for categories of events
with two SS leptons (with the third \PW boson decaying hadronically) and with three leptons.
This search exploits these two event categories.

Certain new physics processes could lead to an excess of events over the SM prediction.
These include, for example, processes with anomalous triple gauge couplings (aTGCs)~\cite{Degrande:2012} and
anomalous QGCs (aQGCs)~\cite{Buchmuller:1985,Grzadkowski:2010es,Degrande:2012,Degrande2014}.
Since this analysis cannot improve the constraints already placed on aTGCs by
recent diboson searches~\cite{Sirunyan:2017bey,Sirunyan:2019bez,Sirunyan:2019dyi,Aad:2014dta,Aad:2016ett,Aaboud:2017cgf},
it focuses on aQGCs.  The production of massive, axionlike particles
(ALPs)~\cite{Peccei:1977hh,Peccei:1977ur,Weinberg:1977ma,Wilczek:1977pj,Graham:2015cka,Brivio:2017ije,Izaguirre:2016dfi,Bauer:2017ris,Dolan:2017osp,Craig:2018kne}
is also considered. In the absence of a signal beyond the SM, limits are placed on
aQGCs and on the production of ALPs in association with \PW bosons.

\section{The CMS detector\label{sec:CMS}}

The central feature of the CMS apparatus is a superconducting solenoid of 6\unit{m} internal diameter,
providing a magnetic field of 3.8\unit{T}. Within the solenoid volume are a silicon pixel and strip tracker, a
lead tungstate crystal electromagnetic calorimeter (ECAL), and a brass and scintillator hadron calorimeter, each composed of
a barrel and two endcap sections. Forward calorimeters extend the pseudorapidity ($\eta$) coverage provided by the
 barrel and endcap detectors. Muons are detected in gas-ionization detectors embedded in the steel flux-return yoke outside
the solenoid.
Events of interest are selected using a two-tiered trigger system~\cite{Khachatryan:2016bia}.
The first level of the CMS trigger system, composed of custom hardware processors, uses information from the
calorimeters and muon detectors to select the most interesting events in a fixed time interval of less than 4\mus. The high-level
 trigger processor farm further decreases the event rate from around 100\unit{kHz} to less than 1\unit{kHz}, before data storage.
A more detailed description of the CMS detector, together with a definition of the coordinate system used and the relevant
kinematic variables, can be found in Ref.~\cite{JINST}.

\section{Data and simulated event samples\label{sec:sim}}

The data are collected using dilepton triggers that select either two electrons, two muons,
or one electron and one muon. These triggers require the leptons to have a high transverse momentum \pt
and to satisfy loose isolation requirements. The dielectron trigger requires
$\pt > 23~(12)\GeV$ for the leading (subleading) electron.  The dimuon trigger requires
$\pt > 17~(8)\GeV$ for the leading (subleading) muon.  Finally, for the electron+muon
trigger, the leading lepton must have $\pt > 23\GeV$ and the subleading lepton must have
$\pt > 12\GeV$ if it is an electron, or $\pt > 8\GeV$ if it is a muon.
Data recorded using prescaled single electron and single muon triggers with \pt thresholds of 8 and 17\GeV, respectively,
are utilized for studies of background rates.
Events with contributions from beam halo processes or anomalous noise in the calorimeter are
rejected using dedicated filters~\cite{Khachatryan:2014gga}.

Samples of simulated events are used to optimize the event selection, to estimate some
of the SM background processes, and to interpret the results in terms of \WWW production.
The \MGvATNLO2.2.2 generator~\cite{Alwall:2014hca} is used
in the next-to-leading-order (NLO) mode with FxFx jet matching~\cite{Frederix:2012ps}
to generate triboson events, both the signal (\WWW including \WH) and the triboson background processes (such as $\PW\PW\PZ$).
The same generator is used in the leading-order (LO) mode with the MLM jet matching~\cite{Alwall:2007fs}
to generate SM \ttbar, \ttbar{}+X ($\mathrm{X} = \PW, \PZ, \PH$), \PW{}+jets,
\PZ{}+jets, \Wg, and \SSWW events.
Other diboson (\WW, \WZ, and \ZZ) events and the single top quark process
are generated at NLO with \POWHEG 2.0~\cite{Nason:2004rx,Frixione:2007vw,Alioli:2010xd,Re:2010bp}.
The most precise cross section calculations available are used to normalize the
simulated samples, and usually correspond to either NLO or next-to-NLO accuracy~\cite{Gavin:2010az,Campbell:2011bn,Czakon:2011xx,Gavin:2012sy,Campbell:2012dh,Garzelli:2012bn,Alwall:2014hca,Cascioli:2014yka,deFlorian:2016spz,Grazzini:2016swo}.

The \MGvATNLO event generator is used in the NLO mode to simulate events
following the model for photophobic, axionlike particles according to the
model described in Ref.~\cite{Craig:2018kne}.
The aQGC samples are generated using \MGvATNLO2.2.2 in the LO mode and the reweighting prescription of Ref.~\cite{Mattelaer:2016}.

The NNPDF3.0~\cite{Ball:2014uwa} parton distribution functions (PDFs) are used for all samples.
Parton showering, hadronization, and the underlying event are modeled by \PYTHIA 8.205~\cite{Sjostrand:2014zea} with parameters set by the CUETP8M1 tune~\cite{Khachatryan:2015pea}.
Additional $\Pp\Pp$ collisions due to multiple interactions in the same or adjacent beam crossings, known as pileup,
are also simulated, and the simulated distribution of pileup interactions is reweighted to match the data.
The response of the CMS detector is simulated with the \GEANTfour~\cite{geant4} package.
The simulated events are reconstructed using the same software as the real data.

\section{Event reconstruction\label{sec:objects}}

The CMS event reconstruction is based on the particle-flow (PF) algorithm~\cite{Sirunyan:2017ulk},  which
combines information from the tracker, calorimeters, and muon systems to identify charged and
neutral hadrons,  photons,  electrons,  and muons, known as PF~candidates.

Each event must contain at least one $\Pp\Pp$ interaction vertex. The reconstructed vertex with the largest value of summed physics-object $\pt^2$ is taken to be the primary vertex (PV). The physics objects are the objects reconstructed by a jet finding algorithm~\cite{Cacciari:2008gp,Cacciari:2005hq,Cacciari:2011ma} applied to all charged particle tracks associated with the vertex and also the corresponding missing transverse momentum (\ptmiss).

Electrons and muons are identified by associating a track reconstructed in the silicon detectors with
either a cluster of energy in the ECAL~\cite{Khachatryan:2015hwa} or a track in the muon system~\cite{Sirunyan:2018fpa}, as appropriate.
To be selected for this analysis, electron and muon candidates must satisfy $\pt > 10\GeV$ and $\abs{\eta} < 2.4$.
Electrons with $1.4 < \abs{\eta} < 1.6$, which corresponds to the transition region
between the barrel and endcap regions of the ECAL, are discarded.
Several working points are defined, which differ according to the identification criteria chosen including the requirements
on the three-dimensional impact parameter \IP\ and relative isolation \IREL.
The impact parameter is the distance between the PV and the point of closest approach of the
lepton track; $\IP < 0.015\unit{cm}$ is required for all lepton candidates. This requirement is tightened to
$\IP < 0.010\unit{cm}$ for electrons in the SS category.
The relative isolation of a lepton with \ptl is defined as
\begin{equation*}
 \IREL  = \left. \left(\sum\ptc + \max\left[\sum\ptnc-\ptPU,0\right]\right) \right/ \ptl.
\end{equation*}
In this expression, $\sum\ptc$ is the scalar \pt sum of charged particles from the PV in a cone of $\Delta R = \sqrt{\smash[b]{(\Delta\eta)^2+(\Delta\phi)^2}} = 0.3$ around the lepton direction, and $\sum\ptnc$ is the equivalent \pt sum for the neutral hadrons and the photons. The lepton momentum itself is not included in  $\sum\ptc$.
The total neutral component contains contributions from pileup, estimated using
$\ptPU=\rho \Aeff$ where the average \pt flow density $\rho$ is calculated in each event
using the jet  area method~\cite{Cacciari:2007fd}, are subtracted.
The effective area \Aeff is the geometric area of the lepton isolation cone multiplied by an $\eta$-dependent factor that accounts for the residual dependence of the isolation on the pileup.
Electrons are required to satisfy $\IREL < 0.03\,(0.05)$ for the SS ($3\ell$) category,
and muons must satisfy $\IREL < 0.03\,(0.07)$.
These leptons are referred to as ``tight'' leptons.
For ``loose'' electrons and muons used in
the estimation of the nonprompt-lepton background, $\IREL < 0.4$ is required.
For ``rejection'' electrons and muons, used to remove background events where extra leptons are present in either the SS or $3\ell$ category,
$\IREL < 0.4$ is required.
For electrons in the SS category, the background contribution coming from a mismeasurement of the track charge is not negligible.
The sign of this charge is inferred using three different observables; requiring all three to agree reduces this background
contribution~\cite{Khachatryan:2015hwa}.

Events containing $\tau$ leptons decaying into charged hadrons are rejected
by requiring no isolated tracks aside from selected electrons and muons.
An isolated track is a charged PF lepton (charged PF hadron) with $\pt > 5~(10)\GeV$,
$\abs{\eta} < 2.4$, and a longitudinal distance to the PV of $\abs{d_z}<0.1\unit{cm}$;
it must be isolated in the sense that $\IREL < 0.2\,(0.1)$  and $\IREL < 8\GeV / \pt^{\text{track}}$.
Any isolated track or lepton that matches a selected lepton candidate within $\Delta R < 0.01$ is discarded.

PF candidates are clustered into jets using the anti-\kt
jet clustering algorithm~\cite{Cacciari:2008gp} with a distance parameter $R=0.4$, implemented in the
\FASTJET package~\cite{Cacciari:2005hq,Cacciari:2011ma}.
Jets must pass loose selection criteria based on the fractions of neutral and charged energy in the jet, and on the relative
amount of electromagnetic and hadronic energy.  Jets with $\pt > 20\GeV$ and $\abs{\eta} < 5$ are selected unless
they are within $\Delta R < 0.4$ of a selected lepton or isolated track.
Jet energies are corrected for contributions from pileup and to account for nonuniform
detector response~\cite{Khachatryan:2016kdb}.
The loose working point of the combined secondary vertex (CSVv2) \PQb~tagging algorithm~\cite{Sirunyan:2017ezt}
is used to identify jets containing the decay of a heavy-flavor hadron.  For this working point, the efficiency
to select \PQb~quark jets is above 80\% and the rate for tagging jets originating
from the hadronization of gluons, and \PQu, \PQd, and \PQs quarks is about 10\%.  In order to apply the CSVv2
\PQb~tagging algorithm, the jet must be reconstructed within $\abs{\eta} < 2.4$.

The vector missing transverse momentum \ptvecmiss is defined as the negative vector \pt sum
of all PF particle candidates. The magnitude of \ptvecmiss is denoted \ptmiss. Corrections to jet energies
due to the nonuniformity in the detector response are propagated to \ptmiss~\cite{Sirunyan:2019kia}.

\section{Search strategy and event selection\label{sec:selection}}

The event selection criteria are designed to maximize the signal significance in the two final states used in the analysis:
two SS leptons and at least two jets (SS~category), and three leptons ($3\ell$~category).
Cross sections for background processes are much larger than the signal cross section,
so stringent requirements must be applied in order to achieve sensitivity to \WWW production.

The SS category contains signal events with the two SS \PW bosons decaying leptonically
and the third \PW boson decaying hadronically. Correspondingly, the selection requires exactly two
tight, high-\pt SS leptons and at least two high-\pt jets.
This category is divided into two signal regions (SRs): ``\Mjj-in'' includes the events in which
the invariant mass of the two jets closest in $\Delta R$ is compatible with the \PW boson mass,
$65 < \Mjj < 95\GeV$; ``\Mjj-out'' includes the remaining events.  The \Mjj-in SR
is expected to contain more signal events and fewer background events than the \Mjj-out region.
The \Mjj-out region still contains a sizable number of \WWW events,
from off-shell \PW bosons from \PW{}\PH production, for example.
It is therefore is considered a signal region.
The main background contribution is called the lost-lepton background and stems from three-lepton events
with one lepton not selected due to an inefficiency (\eg, the isolation requirement)
or because it falls outside the detector acceptance.  Most of this background contribution comes from
\WZ production and a smaller contribution from \ttZ events.  The rejection of events with an extra lepton
or isolated track reduces this background contribution considerably.
A smaller background contribution comes from the production of genuine SS lepton pairs,
mainly through \SSWW{}+ jets and \ttW production.  This contribution
is reduced by requiring the two highest-\pt jets not have a large invariant mass \MjjL or large
$\eta$ separation and by excluding events with \PQb-tagged jets.
Another background contribution comes from events with one or more nonprompt leptons,
such as those from semileptonic decays of heavy-flavor hadrons which arise mainly in \PW{}+jets and
\ttbar{}+jets production. The stringent lepton identification requirements are designed to
suppress this contribution as much as possible.  Additional requirements that \ptmiss be substantial
and that the dilepton mass not be small further suppress
this contribution. In the $\SSem$ channel, a requirement $\MTmax > 90\GeV$
is placed to reduce the contribution from the lost-lepton background from \WZ production; $\MTmax$ is the largest transverse mass
obtained from \ptmiss and any lepton in the event.
Background contributions from events containing misidentified or converted photons and from events
with a lepton charge misassignment are minor.
The details of the event selection for the SS category are listed in Table~\ref{tab:sel:SS}.
There are six SRs defined according to the value of \Mjj (\Mjj-in or \Mjj-out) and
the flavors of the leptons: \SSee, \SSem, or \SSmm.

\begin{table*}[htb]
\topcaption{\label{tab:sel:SS}Event selection criteria for the SS category,
which contains events with two same-sign leptons and at least two hadronic jets.}
\centering
\begin{scotch}{l c c c}
Variable & \SSee & \SSem & \SSmm \\
\hline
Signal leptons & \multicolumn{3}{c}{2 tight same-sign leptons with $\pt>25\GeV$} \\
Additional leptons &  \multicolumn{3}{c}{No additional rejection lepton} \\
Isolated tracks &  \multicolumn{3}{c}{No (additional) isolated tracks} \\
Jets &  \multicolumn{3}{c}{At least two jets with $\pt>30\GeV$, $\abs{\eta}<2.5$} \\
\PQb-tagged jets &  \multicolumn{3}{c}{No \PQb-tagged jet} \\
\multirow{2}{*}{\Mjj (dijet mass of jets closest in $\Delta R$)} &  \multicolumn{3}{c}{$65 < \Mjj < 95\GeV$ (\Mjj-in) OR} \\
 &  \multicolumn{3}{c}{$\abs{\Mjj-80\GeV}\geq15\GeV$ (\Mjj-out)} \\
\MjjL (dijet mass of leading jets) &  \multicolumn{3}{c}{$<$400\GeV} \\
$\Delta\eta$ of two leading jets &  \multicolumn{3}{c}{$<$1.5} \\
\ptmiss &  $>$60\GeV & $>$60\GeV & $>$60\GeV if \Mjj-out \\
\Mll & $>$40\GeV & $>$30\GeV & $>$40\GeV \\
\Mll  & $\abs{\Mll-\MZ} > 10\GeV$ & \NA & \NA \\
\MTmax  & \NA & $>$90\GeV & \NA \\
 \end{scotch}
\end{table*}

The $3\ell$ category contains signal events with all three \PW bosons decaying leptonically, so exactly three charged leptons are required.
The fact that the total charge of the three leptons is $\pm$1 means that there can be zero, one, or two same-flavor, opposite-sign~(SFOS) lepton pairs;
three SRs are designated 0\,SFOS, 1\,SFOS, 2\,SFOS accordingly.  The background sources are similar to those in the SS category.
The contribution from three prompt-lepton final states (mostly \WZ production) is suppressed by requiring
the invariant masses of all SFOS pairs to be incompatible with the \PZ boson mass and with low-mass resonances.
Additional reduction is achieved through the following requirements: if exactly one SFOS lepton pair is found,
the transverse mass \mT calculated from the third lepton and \ptvecmiss, \MTthird,
must be larger than 90\GeV; and, for events with no SFOS pairs, $\MTmax$ is required
to be larger than 90\GeV. These \mT requirements reduce the three-lepton background
contributions, which originate mostly from \WZ production.

Background contributions from nonprompt leptons and converted or misidentified photons are reduced by requiring
large \ptmiss, large \pt of the three-lepton system \ptlll, and a large azimuthal separation
\DPhilllMET between \ptvecmiss and the transverse momentum vector of the three-lepton system, \ptlllvec.
The nonprompt-lepton background from \ttbar production is further reduced by rejecting events
with more than one jet or with any \PQb-tagged jets.  Background contributions from photon conversions
in which the photon is radiated in a \PZ boson decay are suppressed by requiring that the three-lepton
invariant mass \Mlll is not close to the \PZ boson mass.
The details of the $3\ell$ selection requirements are presented in Table~\ref{tab:sel:3l}.

\begin{table*}[htb]
\topcaption{\label{tab:sel:3l}Event selection criteria for the $3\ell$ category,
which contains events with exactly three leptons.}
\centering
\begin{scotch}{lccc}
Variable & 0\,SFOS & 1\,SFOS & 2\,SFOS \\
\hline
\multirow{2}{*}{Signal leptons} & \multicolumn{3}{c}{3 tight leptons with $\pt>25/20/20\GeV$} \\
                                                & \multicolumn{3}{c}{and charge sum = $\pm1$e} \\
Additional leptons &  \multicolumn{3}{c}{No additional rejection lepton} \\
Jets &  \multicolumn{3}{c}{At most one jet with $\pt>30\GeV$, $\abs{\eta}<5$} \\
\PQb-tagged jets &  \multicolumn{3}{c}{No \PQb-tagged jets} \\
\ptlll & \NA & $>$60\GeV & $>$60\GeV \\
\DPhilllMET  & \multicolumn{3}{c}{$>$2.5} \\
\ptmiss  & $>$30\GeV & $>$45\GeV & $>$55\GeV \\
\MTmax &  $>$90\GeV & \NA& \NA \\
\MTthird &  \NA & $>$90\GeV & \NA \\
SF lepton mass & $>$20\GeV & \NA & \NA \\
Dielectron mass  & $\abs{m_{\Pe\Pe}-\MZ} > 15\GeV$ &  \NA &  \NA \\
\multirow{2}{*}{\MSFOS} & \multirow{2}{*}{\NA}  & $\abs{\MSFOS-\MZ}>20\GeV$ & $\abs{\MSFOS-\MZ}>20\GeV$ \\
 &  & and $\MSFOS>20\GeV$ & and $\MSFOS>20\GeV$ \\
\Mlll &  \multicolumn{3}{c}{$\abs{\Mlll-\MZ}>10\GeV$} \\
 \end{scotch}
\end{table*}

For these event selection criteria, about one third of the selected signal events originate
from resonant \PH boson production.

\section{Background estimation\label{sec:bg}}

The background sources for the SS and $3\ell$ categories are essentially the same.
Four such sources are considered: lost leptons, two or three leptons from \PW decays, nonprompt leptons,
and ``other'' minor sources.  The lost-lepton background contributions come from final states with
one or more \PZ bosons: \WZ, \ttZ, and \ZZ.  This contribution is estimated using a three-lepton
control region (CR) with at least one SFOS pair compatible with the decay of a \PZ boson.
The background processes in which the SS lepton pair or all three leptons stem from the decay of a \PW boson, such as from the \ttW process,
are estimated from simulation and validated in an appropriate CR.  Background yields from
nonprompt leptons are calibrated using a CR in which one lepton passes the ``loose''
identification requirements but fails the ``tight'' requirements (as discussed in Section~\ref{sec:objects}).
The other background contributions are predicted using simulated event samples that are validated using the data.
The following sections provide the details of the background estimations.

\subsection{Lost-lepton and three-lepton background\label{sec:bg:WZ}}

The background predictions for both the SS and the $3\ell$ categories rely on the selection
of a pair of leptons consistent with a \PZ boson decay. This background type is expected to contribute
from about one third to over 90\% of the total background yields, depending on the SR.

Simulation suggests that about two thirds of the lost-lepton events in the SRs of the
SS category are present because a lepton does not pass the \pt and $\eta$ requirements.
The remaining lost leptons are rejected by identification and isolation requirements.
For the SS category, events with three leptons are selected. The additional third lepton must have $\pt>20\GeV$.
Among those three leptons, an SFOS lepton pair that satisfies $\abs{\MSFOS-\MZ}<10\GeV$ is required.
All other SS selection criteria listed in Table~\ref{tab:sel:SS} are imposed, except the requirement on \Mjj
is dropped in order to retain a sufficient number of events.
For a given lepton flavor composition (\SSee, \SSem, or \SSmm), the two corresponding SRs of
the \Mjj-in and \Mjj-out selections have one common CR.
In these events, the jets stem from initial-state radiation
and have similar kinematic distributions in both the SRs and CRs, so the extrapolation
from the CR to the SR is reliable.

For the $3\ell$ category, the CRs are defined in a similar fashion.
All selection criteria stated in Table~\ref{tab:sel:3l} are retained, but the requirement $\abs{\MSFOS-\MZ}>20\GeV$
is inverted so that there is at least one SFOS lepton pair compatible with a \PZ boson decay.
Many events are selected for the 1~and 2\,SFOS CRs, but for the 0\,SFOS SR
no corresponding CR exists.  The results are extrapolated from the 1\,SFOS and 2\,SFOS
regions to the 0\,SFOS region as follows:
since the observed and predicted yields agree well in the 1~and 2\,SFOS CRs,
the central value for this background type in the 0\,SFOS SR is taken from simulation, and the
relative systematic uncertainty of the 1\,SFOS SR prediction, as described below, is added to the statistical uncertainty
in the simulated yield.

The transfer factors needed to relate the yields in the CRs to the background contributions
in the SRs are calculated using the simulation.
The observed yields in these CRs agree well with the yields predicted using the simulation.
Corrections to this extrapolation due to differences
between the lepton reconstruction efficiencies in data and simulation are applied, and corresponding uncertainties
are evaluated.
The modeling of the \MSFOS distribution and its associated uncertainty for the SS category is tested using the mass spectrum in the CR. For the $3\ell$ category, in order to ensure no overlap with the SRs, this test is performed after inverting at least one of the SR requirement on \ptmiss, \DPhilllMET, \ptlll, or \MTthird. This validation region has also only a small non-$3\ell$ contamination.
The uncertainty due to limited knowledge of the V\PZ (V = \PW or \PZ) and \ttZ cross sections and their relative contribution in both SRs and CRs is estimated using events from the SS CRs, but after the requirement of no \PQb-tagged jets is removed. The spectrum of the \PQb-tagged jet multiplicity in simulation is fitted to the one observed in data, and the result of that fit is used to assess the uncertainty due to the relative contribution of V\PZ versus \ttZ.
For the SS category, an additional uncertainty due to the \Mjj modeling is evaluated by comparing the
observed and  predicted yields of all CRs. Experimental uncertainties, such as the uncertainty on the
jet energy corrections (JECs), are taken into account.  A correction for the non-$3\ell$
contamination of the CRs is applied. This contamination is small, and stems mostly from nonprompt
leptons or leptons from photon misidentified as electrons. The contamination is estimated from simulation, and a 50\% relative
uncertainty is assigned based on the validation study reported in Section~\ref{sec:bg:minor}.
Uncertainties associated with the CR-to-SR transfer factors are included also.
The impact of all these uncertainties is discussed in Section~\ref{sec:sys}.

A summary of the lost-lepton and three-lepton background estimation is reported in Table~\ref{tab:bg:WZ:pred}.
All CRs are mutually exclusive and do not overlap with any of the SRs.
\begin{table*}[htb]
\topcaption{\label{tab:bg:WZ:pred} Lost-lepton and three-lepton background contributions.
The number of events in the data control regions (CRs) and the non-$3\ell$ contribution,
which are estimated from simulation, are reported together with the control-to-signal region transfer factor
($TF_{\mathrm{CR}\to\mathrm{SR}}$). The predicted background yields obtained from the simulated samples are given as MC prediction. Here, the uncertainty reflects the size of the simulated sample. The last column reports the prediction of the
lost-lepton and three-lepton background contributions to the signal regions,
together with the statistical and systematic uncertainties.}
\centering
   \renewcommand{\arraystretch}{1.2}
     \cmsTable{
\begin{scotch}{l l c c c c c}
\multicolumn{2}{c}{Channel} & Data (CR) & Non-$3\ell$ (CR) & $TF_{\mathrm{CR}\to\mathrm{SR}}$ & MC prediction & Background estimate \\
\hline
\multirow{3}{*}{SS \Mjj-in} & \SSee & 6 & $0.01\pm0.01$ & $0.134^{+0.053}_{-0.066}$ & $0.45\pm0.17$ & $0.80^{+0.48}_{-0.32}\stat^{+0.32}_{-0.40}\syst$  \\
                           & \SSem & 13 & $0.26\pm0.13$ & $0.103^{+0.024}_{-0.024}$ & $1.56\pm0.31$ & $1.31^{+0.48}_{-0.37}\stat^{+0.30}_{-0.30}\syst$  \\
                           & \SSmm & 50 & $1.04\pm0.58$ & $0.062^{+0.011}_{-0.012}$ & $3.04\pm0.48$ & $3.02^{+0.50}_{-0.43}\stat^{+0.54}_{-0.60}\syst$  \\ [\cmsTabSkip]
\multirow{3}{*}{SS \Mjj-out} & \SSee & 6 & $0.01\pm0.01$ & $0.600^{+0.140}_{-0.144}$ & $2.04\pm0.36$ & $3.60^{+2.15}_{-1.43}\stat^{+0.84}_{-0.86}\syst$  \\
                            & \SSem & 13 & $0.26\pm0.13$ & $0.382^{+0.067}_{-0.064}$ & $5.78\pm0.63$ & $4.86^{+1.79}_{-1.36}\stat^{+0.85}_{-0.82}\syst$  \\
                            & \SSmm & 50 & $1.04\pm0.58$ & $0.090^{+0.014}_{-0.014}$ & $4.42\pm0.57$ & $4.39^{+0.73}_{-0.63}\stat^{+0.67}_{-0.68}\syst$  \\ [\cmsTabSkip]
\multirow{3}{*}{3$\ell$} & 0\,SFOS & \NA & \NA & \NA & $0.47\pm0.15$ & $0.47^{+0.20}_{-0.19}\syst$  \\
                           & 1\,SFOS & 34 & $1.01\pm0.53$ & $0.095^{+0.019}_{-0.017}$ & $3.40\pm0.48$  & $3.14^{+0.66}_{-0.55}\stat^{+0.62}_{-0.55}\syst$ \\
                           & 2\,SFOS & 155 & $2.74\pm1.37$ & $0.066^{+0.009}_{-0.009}$ & $10.07\pm0.87$ & $10.10^{+0.89}_{-0.82}\stat^{+1.30}_{-1.30}\syst$ \\
\end{scotch}
}
\end{table*}

\subsection{Background due to nonprompt leptons\label{sec:bg:fake}}

The background contribution from nonprompt leptons is usually relatively small.  However, because of the limited knowledge
of this process, the associated uncertainty can have a significant impact on the result.
The source of this background contribution is \PW{}+jets and \ttbar events in which one or two leptons come from \PW boson decays
and another lepton comes either from a heavy-flavor hadron decay or from misidentified light hadrons.
The background contribution is estimated using the tight-to-loose (TL) method~\cite{Khachatryan:2010ez}.
The implementation used in this analysis is similar to the one used in searches for supersymmetric particles~\cite{Sirunyan:2017uyt}
and accounts for the kinematic properties and flavor of the parent parton of the nonprompt lepton.
The TL method uses two CRs: the measurement region, which is used to extract the TL ratio \eTL;
and the application region (AR), where \eTL is applied to estimate the contribution from the nonprompt-lepton background
to the SRs.
The \eTL measurement region is defined by events containing exactly one loose lepton.
To enrich this region with nonprompt leptons, events with $\ptmiss<20\GeV$ and
$\mT(\ptlvec,\ptvecmiss)<20\GeV$ are selected. To select events with  kinematic properties
 similar to those in \PW{}+jets and \ttbar events,
the presence of at least one jet with $\pt>40\GeV$, $\abs{\eta}<2.4$ and $\Delta R(\ptlvec,\ptjetvec)>1$ is required.
The TL ratio is defined as the fraction of events in the measurement region in which the loose lepton
also passes the tight lepton selection; and \eTL is computed as a function of \ptcor and $\abs{\eta}$.
Here, \ptcor is \ptl plus the fraction of the \pt sum of objects in the isolation cone exceeding
the isolation threshold value defined in Section~\ref{sec:objects}.  The quantity \ptcor is better
correlated with the parent parton \pt than is \ptl.
The \eTL measurement is corrected for the contribution of prompt leptons in the measurement region.
This contribution is taken from simulation, but its normalization is taken from data in the measurement region
sideband satisfying $\ptmiss>30\GeV$ and $80<\mT(\ptlvec,\ptvecmiss)<120\GeV$. Uncertainties in the
extrapolation from the sideband to the measurement region are evaluated; they are dominated by the JEC uncertainty.

The ARs are defined similarly to the SRs, with the difference that one of the leptons only passes the loose
but not the tight selection defined in Section~\ref{sec:objects}.  Nonprompt leptons are the main contribution
to these regions; small contributions from prompt lepton events are estimated with simulations and subtracted.
The background contribution is estimated by weighting each event by $\eTL/(1-\eTL)$, where \eTL is
the probability that the lepton fails the tight selection, and summing all the event weights.

The performance of the TL method is evaluated in simulation by comparing the prediction of the TL method in the SR
with the actual yield of nonprompt-lepton background; they agree within the statistical precision of this test.
The statistical uncertainty of the test is assigned as an additional systematic uncertainty.
The results of the nonprompt-lepton background estimation with its systematic uncertainties are
given in Table~\ref{tab:bg:fake:results}.

\begin{table*}[htb]
    \topcaption{\label{tab:bg:fake:results} Nonprompt-lepton background estimates. The data in the application regions (AR),
    the prompt yields (AR) from simulations,  and the predicted nonprompt-lepton background are reported.
    The uncertainties in the prediction are split into statistical and systematic components.}
\centering
\begin{scotch}{l c c c c }
\multicolumn{2}{c}{Channel}     & Data (AR) & Prompt yield (AR) & Background estimate                                   \\
\hline
\multirow{3}{*}{SS \Mjj-in}     & \SSee     & 8         & $3.2\pm2.2$ & $0.89 \pm 0.53\stat \pm 0.63\syst$ \\
                                            & \SSem    & 16       & $1.7\pm0.3$ & $0.92 \pm 0.26\stat \pm 0.43\syst$ \\
                                            & \SSmm   & 57       & $2.9\pm0.5$ & $0.82 \pm 0.11\stat \pm 0.36\syst$ \\ [\cmsTabSkip]
\multirow{3}{*}{SS \Mjj-out}  & \SSee     & 4         & $1.1\pm0.5$ & $0.47 \pm 0.32\stat \pm 0.28\syst$ \\
                                            & \SSem    & 32       & $2.8\pm0.5$ & $1.60 \pm 0.31\stat \pm 0.64\syst$ \\
                                            & \SSmm   & 36       & $3.2\pm0.5$ & $0.59 \pm 0.11\stat \pm 0.25\syst$ \\ [\cmsTabSkip]
\multirow{3}{*}{$3\ell$}        & 0\,SFOS  & 17       & $0.7\pm0.3$ & $0.97 \pm 0.25\stat \pm 0.22\syst$ \\
                                            & 1\,SFOS  & 2         & $0.8\pm0.3$ & $0.07^{+0.08}_{-0.07}\stat^{+0.11}_{-0.07}\syst$ \\
                                            & 2\,SFOS  & 6         & $2.0\pm0.5$ & $0.30 \pm 0.18\stat \pm 0.25\syst$ \\
\end{scotch}
\end{table*}

\subsection{Irreducible backgrounds\label{sec:bg:WW:WWW}}

The third important background process for this search is irreducible, namely, two or three charged leptons originating from \PW boson decays.
This background process is similar to the signal process and is estimated using Monte Carlo simulations.  For the SS category, the simulation predicts
that 49\% of this background process comes from \ttV production (mostly \ttW), 47\% from \SSWW{}+ jets, and 4\% from double-parton scattering (DPS) \SSWW.
For the $3\ell$ category, the irreducible background process comes almost completely from  \ttW production.
The uncertainty for this background process is based on the relevant cross section measurements by the CMS Collaboration:
for \ttW production the uncertainty is 22\%~\cite{Sirunyan:2017uzs} and for \SSWW{}+ jets it is 20\%~\cite{Sirunyan:2017ret}.
The estimation of this background process is verified in certain validation regions in which the dominant contribution
comes from the \ttW\ process.  The validation regions, however, are not as pure as those  defined for the lost-lepton or
nonprompt-lepton backgrounds.
For the \ttW contribution, the validation region is defined by requiring events to contain two tight SS leptons, $\geq$4 jets, $\geq$1 \PQb-tagged jets
and  $60<\Mjj<100\GeV$. For the \SSWW{}+ jets contribution, the validation region is constructed by requiring two tight SS leptons,
$\geq$2 jets, 0\,\PQb-tagged jets, $\MjjL>400\GeV$, and $\abs{\DetaJJ}>1.5$.
The observed yields and the estimates based on simulations agree within the statistical power of the test.

\subsection{Other backgrounds\label{sec:bg:minor}}

Other remaining background yields are expected to be very small.
They originate from either a charge misassignment for one of the leptons or from events containing a photon that is either misidentified
as an electron, or that converts to an $\ell^{+}\ell^{-}$ pair with one of the leptons being lost.
These contributions are estimated using simulation and are validated with data.
The background yields due to lepton charge misassignment are validated in a
dielectron sample with $\abs{\Mll-\MZ}<10\GeV$ by comparing the events yields when the two electrons have either the equal or opposite electric charge.
The background contribution due to events with leptons originating from photons is validated in a three-lepton validation region enriched in $\PZ\gamma$ production.
The selection is similar to the $3\ell$ SR selection (Table~\ref{tab:sel:3l}), but at least one SFOS lepton pair with
$\abs{\MSFOS-\MZ}<20\GeV$ is required. Also the requirement on \Mlll is dropped and the one on \ptlll is inverted.
A 50\% relative uncertainty is assigned to these background sources.
Within this uncertainty, the agreement between data and simulation in these validation regions is satisfactory.

\section{Systematic uncertainties\label{sec:sys}}

The systematic uncertainties of the estimated background contributions are discussed in Section~\ref{sec:bg} and
a detailed summary is provided in Table~\ref{tab:sys:bg}.
Systematic uncertainties associated with the \WWW event production are described below and are summarized in Table~\ref{tab:sys:sig}.

\begin{table*}[htb]
  \topcaption{\label{tab:sys:bg} Summary of typical systematic uncertainties in estimated background contributions.
    The ranges indicate variations across different signal regions.}
  \centering
  \begin{scotch}{l c c c c c}
    \multirow{2}{*}{Uncertainty}        & Lost-lepton/   & Nonprompt & \multirow{2}{*}{$\gamma\to\ell$} & Charge mis- & Irreduc- \\
                                                     & three-lepton & leptons        &    & assignment  & ible \\
\hline
 Control data sample size                 & 11--46\%   & 15--43\%   & \NA        & \NA       & \NA \\
 Simulation statistical uncertainty     & 14--25\%    &  \NA             & \NA        & \NA       & 4--18\% \\
 Lepton reconstruction                   & $<$1\%      &    \NA           & \NA        & \NA       & $<$1\% \\
 Lepton energy resolution               & $<$1\%      &   $<$1\%   & \NA       & \NA       & $<$1\% \\
 \Mjj modeling (SS only)                & 7.3\%         & \NA              & \NA        & \NA       & \NA \\
 Jet energy scale                       & 1--7\%       & \NA              & \NA        & \NA       & \NA \\
 \MSFOS extrapolation                   & 5-8\%        & \NA             & \NA        & \NA       & \NA \\
 \ttZ/\WZ fraction                 & $<$1\%      & \NA              & \NA        & \NA       & \NA \\
 \eTL measurement                      & \NA               & 21--43\%   & \NA        & \NA       & \NA \\
 Validation of TL ratio method     & \NA               & 22--25\%  & \NA        & \NA       & \NA \\
 \PQb tagging                             & $<$1\%      & \NA             & \NA        & \NA       & 2--4\% \\
 Cross section measurement         & \NA              & \NA             & \NA        & \NA       & 14--22\% \\
 Trigger                                           & \NA              & \NA             & \NA        & \NA       & 1\% \\
 Pileup                                           & 1--8\%        & \NA             & \NA        & \NA       & \NA \\
 Integrated luminosity                    & \NA              & \NA             & \NA        & \NA       & 2.5\% \\
 Other uncertainties                        & \NA              & \NA             & 50\% & 50\% & \NA \\
\end{scotch}
\end{table*}

The experimental uncertainties for the signal include JECs~\cite{Khachatryan:2016kdb,CMS-DP-2016-020},
lepton energy resolution, lepton efficiency data-to-simulation correction factors \cite{Khachatryan:2015hwa,Sirunyan:2018fpa},
\PQb tagging correction factors~\cite{Sirunyan:2017ezt}, trigger efficiencies, pileup, and integrated luminosity~\cite{CMS:2017sdi}
uncertainties. The lepton reconstruction efficiencies and trigger efficiencies are measured with a tag-and-probe
method~\cite{CMS:2011aa} applied to $\PZ\to\ell^{+}\ell^{-}$ events.

The theoretical uncertainty for the predicted signal cross section is obtained from  Ref.~\cite{Dittmaier:2017bnh}.
Uncertainties in the signal acceptance from the renormalization~($\muR$) and factorization~($\muF$)
scales are evaluated~\cite{Catani:2003zt,Cacciari:2003fi,Kalogeropoulos:2018cke}.
Parametric (PDF and \alpS) uncertainties are estimated using the PDF4LHC prescription~\cite{Butterworth:2015oua} with the NNPDF3.0 set~\cite{Ball:2014uwa}.
The impact of the systematic uncertainties on the signal is small compared to those of the background estimations.

\begin{table}[hb]
  \topcaption{\label{tab:sys:sig} Summary of systematic uncertainties for the signal process.}
  \centering
  \begin{scotch}{l c}
    Uncertainty & Typical size \\
    \hline
    Simulation statistical uncertainty                               & 12--33\%  \\
    Cross section calculation (normalization)                   & 6\%         \\
    \muR/\muF (acceptance only)                                   & 1--13\%    \\
    PDF (acceptance only)                                              & 1--4\%      \\
    \alpS                                                                          & 1\%           \\
    Lepton reconstruction efficiency                                & 2--3\%      \\
    Lepton energy resolution                                           & 0--2\%      \\
    Jet energy scale                                                        & 1--7\%      \\
    \PQb tagging scale factor                                         & 1--3\%       \\
    Trigger                                                                       & 3--5\%      \\
    Pileup                                                                        & 0--4\%      \\
    Luminosity                                                                 & 2.5\%       \\
  \end{scotch}
\end{table}

\section{Results and interpretations\label{sec:result}}
\par
This section firstly presents the event yields in the nine nonoverlapping categories used to obtain
the measured value of the production cross section.  Secondly, contributions
to the yield originating from  aQGCs are considered.
Finally, a possible signal from a specific beyond-the-SM model,
photophobic axionlike particle production~\cite{Craig:2018kne}, is investigated.

\subsection{Cross section measurement}
\label{sec:crosssection}

\begin{figure*}[hbt]
\centering
\includegraphics[width=0.75\textwidth]{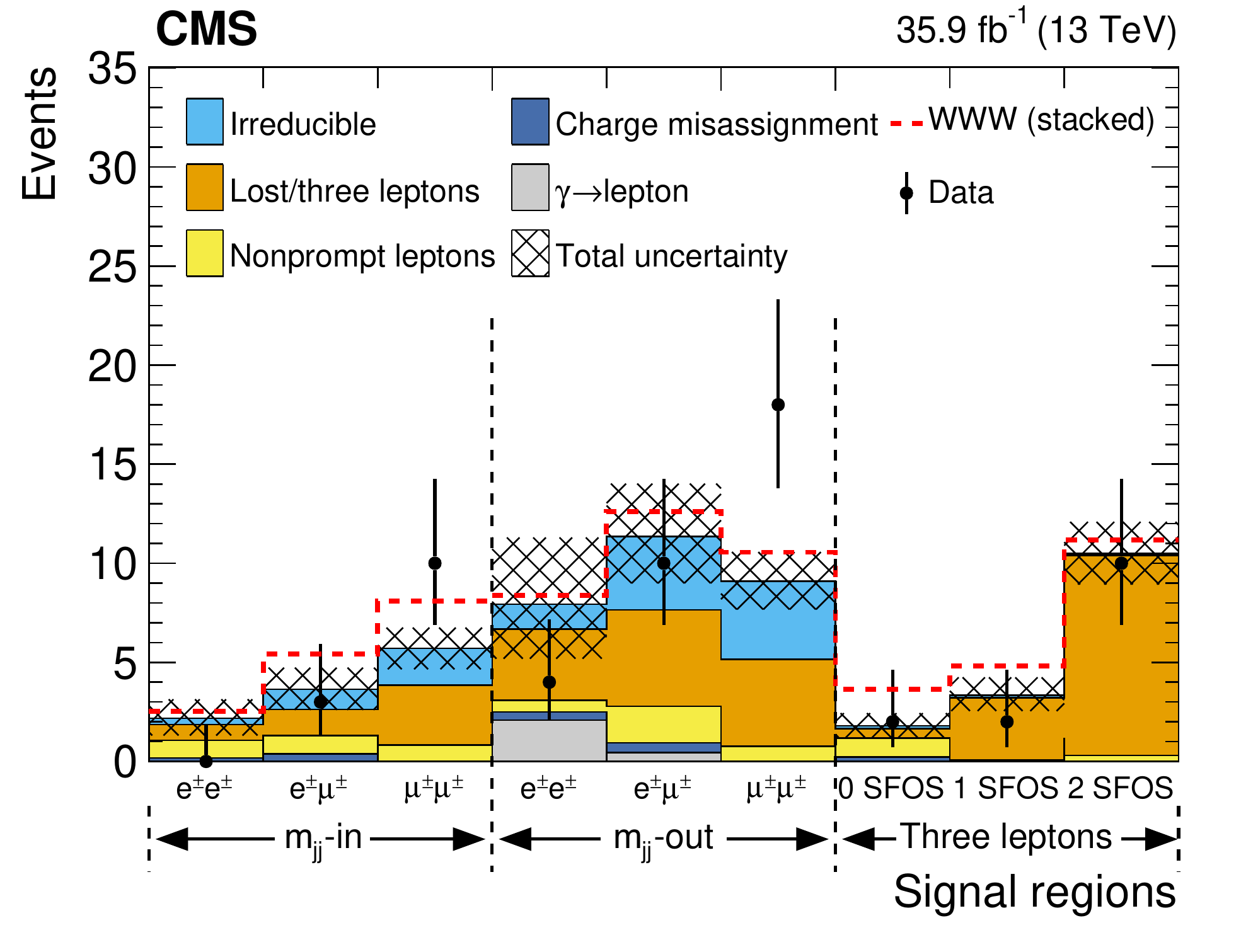}
\caption{\label{fig:results} Comparison of the observed numbers of events to the
predicted yields in the nine signal regions.  The \WWW signal shown is stacked
on top of the total background and is based on the SM theoretical cross section.}
\end{figure*}

The data in all SRs, together with the predicted background yields and expected signal yields, are provided in
Table~\ref{tab:results}.  The \WHtoWWW process contributes  about one third of the expected signal yield.
A graphical representation is given in Fig.~\ref{fig:results}.

\begin{table*}[htb]
\centering
\topcaption{\label{tab:results} Numbers of observed events for all signal regions,
including predicted background contributions and expected signal yields.
The uncertainties presented include both the statistical and systematic uncertainties.}
   \renewcommand{\arraystretch}{1.2}
  \cmsTable{
\begin{scotch}{lccccccccc}
                                & \multicolumn{3}{c}{\Mjj-in} & \multicolumn{3}{c}{\Mjj-out} & \multicolumn{3}{c}{3$\ell$} \\
                                & \SSee & \SSem & \SSmm & \SSee & \SSem & \SSmm & 0\,SFOS & 1\,SFOS & 2\,SFOS \\
\hline
Lost/three $\ell$            & $ 0.8^{+ 0.6}_{- 0.5}$ & $ 1.3^{+ 0.6}_{- 0.5}$ & $ 3.0^{+ 0.7}_{- 0.7}$ & $ 3.6^{+ 2.3}_{- 1.6}$ & $ 4.9^{+ 1.9}_{- 1.5}$ & $ 4.4^{+ 0.9}_{- 0.9}$ & $ 0.5^{+ 0.2}_{- 0.2}$ & $ 3.1^{+ 0.8}_{- 0.7}$ & $10.1^{+ 1.3}_{- 1.2}$ \\
Irreducible                  & $ 0.3^{+ 0.1}_{- 0.1}$ & $ 1.0^{+ 0.2}_{- 0.2}$ & $ 1.9^{+ 0.3}_{- 0.3}$ & $ 1.3^{+ 0.2}_{- 0.2}$ & $ 3.7^{+ 0.4}_{- 0.4}$ & $ 3.9^{+ 0.4}_{- 0.4}$ & $ 0.2^{+ 0.0}_{- 0.0}$ & $ 0.1^{+ 0.1}_{- 0.1}$ & $ 0.1^{+ 0.1}_{- 0.1}$ \\
Nonprompt $\ell$            & $ 0.9^{+ 0.7}_{- 0.7}$ & $ 0.9^{+ 0.8}_{- 0.8}$ & $ 0.8^{+ 0.6}_{- 0.6}$ & $ 0.6^{+ 0.6}_{- 0.5}$ & $ 1.8^{+ 1.4}_{- 1.4}$ & $ 0.8^{+ 0.5}_{- 0.5}$ & $ 1.0^{+ 0.6}_{- 0.5}$ & $ 0.1^{+ 0.1}_{- 0.1}$ & $ 0.3^{+ 0.2}_{- 0.2}$ \\
Charge flips                 & $ 0.2^{+ 0.2}_{- 0.2}$ & $ 0.4^{+ 0.3}_{- 0.2}$ & $<$0.1 & $ 0.4^{+ 0.3}_{- 0.3}$ & $ 0.5^{+ 0.3}_{- 0.3}$ & $<$0.1 & $ 0.2^{+ 0.1}_{- 0.1}$ & $<$0.1 & $<$0.1 \\
$\gamma\to$ nonprompt $\ell$ & $ 0.2^{+ 0.1}_{- 0.1}$ & $ 0.1^{+ 0.1}_{- 0.1}$ & $<$0.1 & $ 2.2^{+ 2.1}_{- 2.1}$ & $ 0.4^{+ 0.5}_{- 0.4}$ & $<$0.1 & $<$0.1 & $<$0.1 & $<$0.1 \\
Background sum               & $ 2.4^{+ 1.0}_{- 0.8}$ & $ 3.7^{+ 1.1}_{- 1.0}$ & $ 5.6^{+ 1.0}_{- 1.0}$ & $ 8.1^{+ 3.2}_{- 2.8}$ & $11.3^{+ 2.5}_{- 2.2}$ & $ 9.1^{+ 1.2}_{- 1.1}$ & $ 1.8^{+ 0.6}_{- 0.6}$ & $ 3.3^{+ 0.8}_{- 0.7}$ & $10.4^{+ 1.3}_{- 1.2}$ \\ [\cmsTabSkip]
\WWW signal                      & $0.3_{-0.1}^{+0.1}$ & $1.8_{-0.3}^{+0.3}$ & $2.4_{-0.3}^{+0.3}$ & $0.4_{-0.2}^{+0.2}$ & $1.3_{-0.3}^{+0.3}$ & $1.5_{-0.4}^{+0.4}$ & $1.8_{-0.4}^{+0.4}$ & $1.5_{-0.3}^{+0.3}$ & $0.7_{-0.3}^{+0.3}$\\
Total               & $ 2.7^{+ 1.0}_{- 0.8}$ & $ 5.5^{+ 1.1}_{- 1.0}$ & $ 7.9^{+ 1.0}_{- 1.0}$ & $ 8.5^{+ 3.2}_{- 2.7}$ & $12.6^{+ 2.5}_{- 2.2}$ & $ 10.6^{+ 1.3}_{- 1.2}$ & $ 3.6^{+ 0.7}_{- 0.7}$ & $ 4.8^{+ 0.9}_{- 0.8}$ & $11.1^{+ 1.3}_{- 1.2}$ \\
Observed                    & $0$ & $3$ & $10$ & $4$ & $10$ & $18$ & $2$ & $2$ & $10$\\
\end{scotch}
}
\end{table*}

A profile maximum likelihood method is used following the procedures set by the LHC Higgs Combination Group~\cite{CMS-NOTE-2011-005}
to extract the expected and observed significances of this analysis to the SM \WWW production process.
The signal strength is constrained to be non-negative.
The systematic uncertainties are treated as nuisance parameters and are profiled in the maximum likelihood fit.
Using the significance as metric, the most sensitive categories among those shown in
Fig.~\ref{fig:results} are 0\,SFOS, \Mjj-in \SSem, 1\,SFOS, and \Mjj-in \SSmm.
For quantifying the absence of a signal, the modified frequentist \CLs statistic~\cite{Junk:1999kv,Read:2002hq}
is used and asymptotic formulae~\cite{Cowan:2010js} are used for quantifying the significance of an excess.

The expected significance for the combined SS and $3\ell$ categories
is 1.78\,standard deviations (s.d.) assuming the SM production of \WWW events, whereas the
observed significance is 0.60\unit{s.d.}
The corresponding expected and observed $p$-values for the null hypothesis are $0.038$ and $0.274$.
The best fit for the observed signal strength, defined as the
ratio of the observed signal to the theoretically predicted one, is $0.34^{+0.62}_{-0.34}$.
It follows that the measured cross section is
\begin{equation*}
  \sigma(\Pp\Pp\to\WWWpm) = 0.17_{-0.17}^{+0.32}  \unit{pb} .
\end{equation*}
The uncertainties include both statistical and systematic components.
Assuming the presence of background only,
the observed (expected) 95\% confidence level (\CL) upper limit on the cross section is
$0.78$\,($0.60)\unit{pb}$.

\subsection{Limits on anomalous quartic gauge couplings \label{sec:aQGC}}

The interaction of four gauge bosons depicted in Fig.~\ref{fig:WWW:feynman}
exists in the SM and contributes to the production
of the \WWW final state.  New physics beyond the SM could be manifested as
an apparent change in the coupling constant associated with the four-boson vertex,
\ie, in an aQGC.  A description based on
aQGCs is appropriate when the mass scale for new physics $\Lambda$ is much higher
than the energy scale of the given process, in this case, \WWW production
characterized by the squared invariant mass of the three \PW bosons, \hatsWWW.

Anomalous couplings can be handled theoretically by extending the SM Lagrangian
with the operator product expansion~\cite{Degrande2014}:
\begin{equation*}
 {\mathcal{L}} =  {\mathcal{L}}_{\textrm{SM}} +
     \sum_{i}\frac{c_{i}}{\Lambda^{2}} {\mathcal{O}}_{i} +
     \sum_{j}\frac{f_{j}}{\Lambda^{4}} {\mathcal{O}}_{j} + \cdots ,
\label{eqn:eft}
\end{equation*}
where ${\mathcal{O}}$ represents the higher-order dimension-6 and dimension-8 operators with
Wilson coefficients $c_{i}$ and $f_{j}$, respectively.
The operators ${\mathcal{O}}_i$ are constructed from SM fields and respect gauge invariance.
The coefficients are unknown and are treated as free parameters to be determined by the data.
The coefficients for all dimension-6 operators, which represent aTGCs, are taken to be zero.
The following dimension-8, CP-conserving operators can be included in the
non-SM part of the Lagrangian~\cite{PhysRevD.74.073005,Degrande2014}:

\begin{linenomath}
\ifthenelse{\boolean{cms@external}}
{
{
\belowdisplayskip=0cm
\abovedisplayskip=0cm
\begin{align*}
 {\mathcal{O}}_{{\textrm{S,0}}}  &=   \left[ (D_{\mu} \Phi)^{\dagger} D_{\nu} \Phi \right]  \left[ (D^{\mu} \Phi)^{\dagger} D^{\nu} \Phi \right],  \\
 {\mathcal{O}}_{{\textrm{S,1}}}  &=   \left[ (D_{\mu} \Phi)^{\dagger} D^{\mu} \Phi \right]  \left[ (D_{\nu} \Phi)^{\dagger} D^{\nu} \Phi \right] ,  \\
\end{align*}
\begin{align*}
 {\mathcal{O}}_{{\textrm{M,0}}}  &=  Tr\left[ \hat{W}_{\mu\nu}  \hat{W}^{\mu\nu} \right]  \left[ (D_{\beta} \Phi)^{\dagger} D^{\beta} \Phi \right], \\
 {\mathcal{O}}_{{\textrm{M,1}}}  &=  Tr\left[ \hat{W}_{\mu\nu}  \hat{W}^{\nu\beta} \right]  \left[ (D_{\beta} \Phi)^{\dagger} D^{\mu} \Phi \right], \\
 {\mathcal{O}}_{{\textrm{M,6}}}  &=  \left[ (D_{\mu} \Phi)^{\dagger} \hat{W}_{\beta\nu} \hat{W}^{\beta\nu}  D^{\mu} \Phi\right],  \\
 {\mathcal{O}}_{{\textrm{M,7}}}  &=  \left[ (D_{\mu} \Phi)^{\dagger} \hat{W}_{\beta\nu} \hat{W}^{\beta\mu}  D^{\nu} \Phi\right],
\end{align*}
\begin{align*}
 {\mathcal{O}}_{{\textrm{T,0}}}  &=   Tr\left[W_{\mu\nu}  W^{\mu\nu} \right]  Tr\left[ W_{\alpha\beta}  W^{\alpha\beta} \right], \\
 {\mathcal{O}}_{{\textrm{T,1}}}  &=   Tr\left[W_{\alpha\nu}  W^{\mu\beta} \right]  Tr\left[ W_{\mu\beta}  W^{\alpha\nu} \right], \\
 {\mathcal{O}}_{{\textrm{T,2}}}  &=   Tr\left[W_{\alpha\mu}  W^{\mu\beta} \right]  Tr\left[ W_{\beta\nu}  W^{\nu\alpha} \right].
 \end{align*}
 }
}
{
\begin{center}
\begin{tabular}{ l l }
 ${\mathcal{O}}_{{\textrm{S,0}}}  =   \left[ (D_{\mu} \Phi)^{\dagger} D_{\nu} \Phi \right]  \left[ (D^{\mu} \Phi)^{\dagger} D^{\nu} \Phi \right]$, &
 ${\mathcal{O}}_{{\textrm{S,1}}}  =   \left[ (D_{\mu} \Phi)^{\dagger} D^{\mu} \Phi \right]  \left[ (D_{\nu} \Phi)^{\dagger} D^{\nu} \Phi \right]$, \\
 ${\mathcal{O}}_{{\textrm{M,0}}}  =  Tr\left[ \hat{W}_{\mu\nu}  \hat{W}^{\mu\nu} \right]  \left[ (D_{\beta} \Phi)^{\dagger} D^{\beta} \Phi \right]$, &
 ${\mathcal{O}}_{{\textrm{M,1}}}  =  Tr\left[ \hat{W}_{\mu\nu}  \hat{W}^{\nu\beta} \right]  \left[ (D_{\beta} \Phi)^{\dagger} D^{\mu} \Phi \right]$, \\
 ${\mathcal{O}}_{{\textrm{M,6}}}  =  \left[ (D_{\mu} \Phi)^{\dagger} \hat{W}_{\beta\nu} \hat{W}^{\beta\nu}  D^{\mu} \Phi\right]$, &
 ${\mathcal{O}}_{{\textrm{M,7}}}  =  \left[ (D_{\mu} \Phi)^{\dagger} \hat{W}_{\beta\nu} \hat{W}^{\beta\mu}  D^{\nu} \Phi\right]$, \\
 ${\mathcal{O}}_{{\textrm{T,0}}}  =   Tr\left[W_{\mu\nu}  W^{\mu\nu} \right]  Tr\left[ W_{\alpha\beta}  W^{\alpha\beta} \right]$, &
 ${\mathcal{O}}_{{\textrm{T,1}}}  =   Tr\left[W_{\alpha\nu}  W^{\mu\beta} \right]  Tr\left[ W_{\mu\beta}  W^{\alpha\nu} \right]$, \\
 ${\mathcal{O}}_{{\textrm{T,2}}}  =   Tr\left[W_{\alpha\mu}  W^{\mu\beta} \right]  Tr\left[ W_{\beta\nu}  W^{\nu\alpha} \right]$. & \\
\end{tabular}
\end{center}
}
\end{linenomath}

The Lagrangian including dimension-8 anomalous coupling terms is:

\begin{align*}
    {\mathcal{L}} = {\mathcal{L}}_{{\textrm{SM}}} +
    & \LTERM{S,0} + \LTERM{S,1} + \LTERM{M,0} +\\
    & \LTERM{M,1} + \LTERM{M,6} + \LTERM{M,7} + \\
    & \LTERM{T,0} + \LTERM{T,1} + \LTERM{T,2} ,
\end{align*}
where the coefficients $f_{x,n} / \Lambda^4$ have dimension \TeV$^{-4}$.
No form factors for enforcing unitarity are employed in this analysis.
When looking for evidence of anomalous couplings, \WWW production as
predicted in the SM is taken as a background process.
Interference effects between the SM and the anomalous contribution to
\WWW production are taken into account.

Since \hatsWWW cannot be measured directly, the kinematic quantity \ST is employed, which is
the sum of the \pt of the leptons and the jets, and \ptmiss.  The presence of aQGCs
would be manifested as an excess of events at high \ST.  Since non-\WWW background
events and SM \WWW events appear at low \ST, a requirement of
$\ST > \STmin$ is imposed.  The value for \STmin is chosen to
optimize the expected limits on the anomalous coupling $\fTa / \Lambda^4$ for which this analysis
is most sensitive.  For the SS and $3\ell$ categories, the values are $\STmin = 2.0$ and~1.5\TeV, respectively.
There is little sensitivity to the operators involving Higgs doublet terms.

The event selection is the same as described in Section~\ref{sec:selection}, except
that the restriction $\MjjL<400\GeV$ on the invariant mass of the leading two jets is removed to retain
sensitivity to aQGCs.
All SRs of the SS category (Table~\ref{tab:sel:SS}) and the $3\ell$ category (Table~\ref{tab:sel:3l}) are merged into one SS and one $3\ell$ SR, respectively.
After the \ST requirement stated above, the numbers of events expected in the SM are very small:
$0.22 \pm 0.10$ events in the SS category (mainly \SSWW{}+jets events) and
less than 0.01~event in the $3\ell$ category.
The systematic uncertainty assigned to the
predicted background yields is 30\% but the predicted limits on anomalous couplings
are insensitive to this uncertainty.
Furthermore, higher-order corrections might reduce the production cross section~\cite{Schonherr:2018}. As a test the signal yield was reduced by 25\% and it was found that the allowed range of anomalous couplings was increased by about 11\%.

No events are selected when the event selection criteria are imposed on the data.
In the absence of any indication for anomalous couplings, limits are set as
summarized in Table~\ref{tab:aQGC_results}.  When calculating the limit on one
anomalous coupling, the others are taken to be zero.

\begin{table}[htb]
\centering
\topcaption{\label{tab:aQGC_results}
Limits on three anomalous quartic couplings at 95\% \CL.}
\begin{scotch}{ c  c c  }
Anomalous coupling & \multicolumn{2}{ c }{Allowed range  (${\TeVns}^{-4}$) } \\
 & Expected & Observed \\
\hline
$\fTa/\Lambda^4$ & [-1.3, 1.3] & [-1.2, 1.2] \\
$\fTb/\Lambda^4$ & [-3.7, 3.7] & [-3.3, 3.3] \\
$\fTc/\Lambda^4$ & [-3.0, 2.9] & [-2.7, 2.6] \\
\end{scotch}
\end{table}

\subsection{Limits on photophobic axionlike particle models \label{sec:alpsearch}}

Since the discovery of a \PH boson~\cite{Aad:2012tfa,Chatrchyan:2012xdj,Chatrchyan:2013lba}, searches for extended scalar sectors
have been of high interest~\cite{Azevedo:2018llq,Branco:2011iw}.  For example, pseudoscalar particles like the quantum chromodynamics axion, which
solve the strong CP problem~\cite{Peccei:1977hh,Peccei:1977ur,Weinberg:1977ma,Wilczek:1977pj}, can also be candidates for dark matter \cite{Preskill:1982cy,Abbott:1982af,Dine:1982ah}.
Other examples address the hierarchy problem via relaxation mechanisms through the relaxion field~\cite{Hook:2016mqo}.
An ALP can have a variety of couplings to SM gauge bosons. Recently, theoretical studies have been extended to include couplings
to gauge bosons besides photons~\cite{Brivio:2017ije,Izaguirre:2016dfi,Bauer:2017ris,Dolan:2017osp}.
Generally speaking, if the ALPs are sufficiently light, branching fractions to photons are expected to be large.

In this study, photophobic ALPs~\cite{Craig:2018kne} are considered whose mass is large enough that their
dominant decay mode is $\Palp \to \WW$. In this scenario, the
\WWW final state results from the production of $\PW {\Palp}$ followed by ${\Palp} \to \WW$.
The \WWW channel has the largest product of production cross section and branching fraction
for $\malp\gtrsim2 \mW$,~\cite{Craig:2018kne}. For $\malp \lesssim 2 \mW$, the
branching fraction falls off rapidly;
the interpretation for $\malp < 200\GeV$ is left for future analyses.
The model has one free parameter, $1/\falp$, which fully determines the
couplings of the ALP of mass $\malp$ to SM particles. In this context, as for aQGCs discussed
in Section~\ref{sec:aQGC}, the SM production of \WWW is treated as a
background to new physics.

For the ALP interpretation, the nine SRs developed for the SM analysis
(Tables~\ref{tab:sel:SS} and~\ref{tab:sel:3l}) are used.
The acceptance of the model in these SRs follows an expected pattern:
when $\malp=200\GeV$, the acceptance is similar to that estimated for the SM \WWW signal process.
As $\malp$ increases, the acceptance rises because the events are more centrally produced
and the decay products more often fall within the fiducial region.

\begin{figure}[htb]
    \centering
    \includegraphics[width=0.49\textwidth]{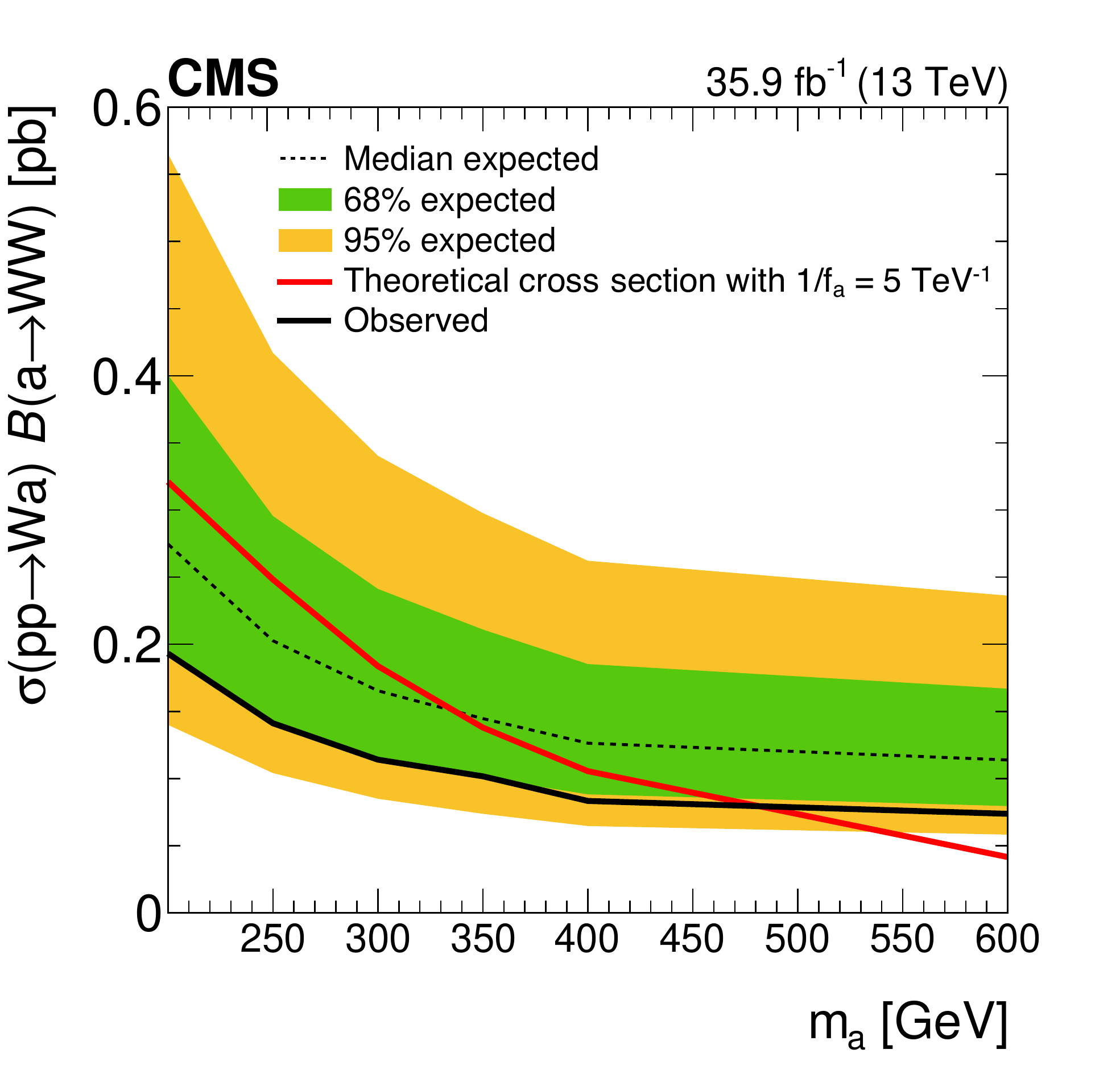}
    \includegraphics[width=0.49\textwidth]{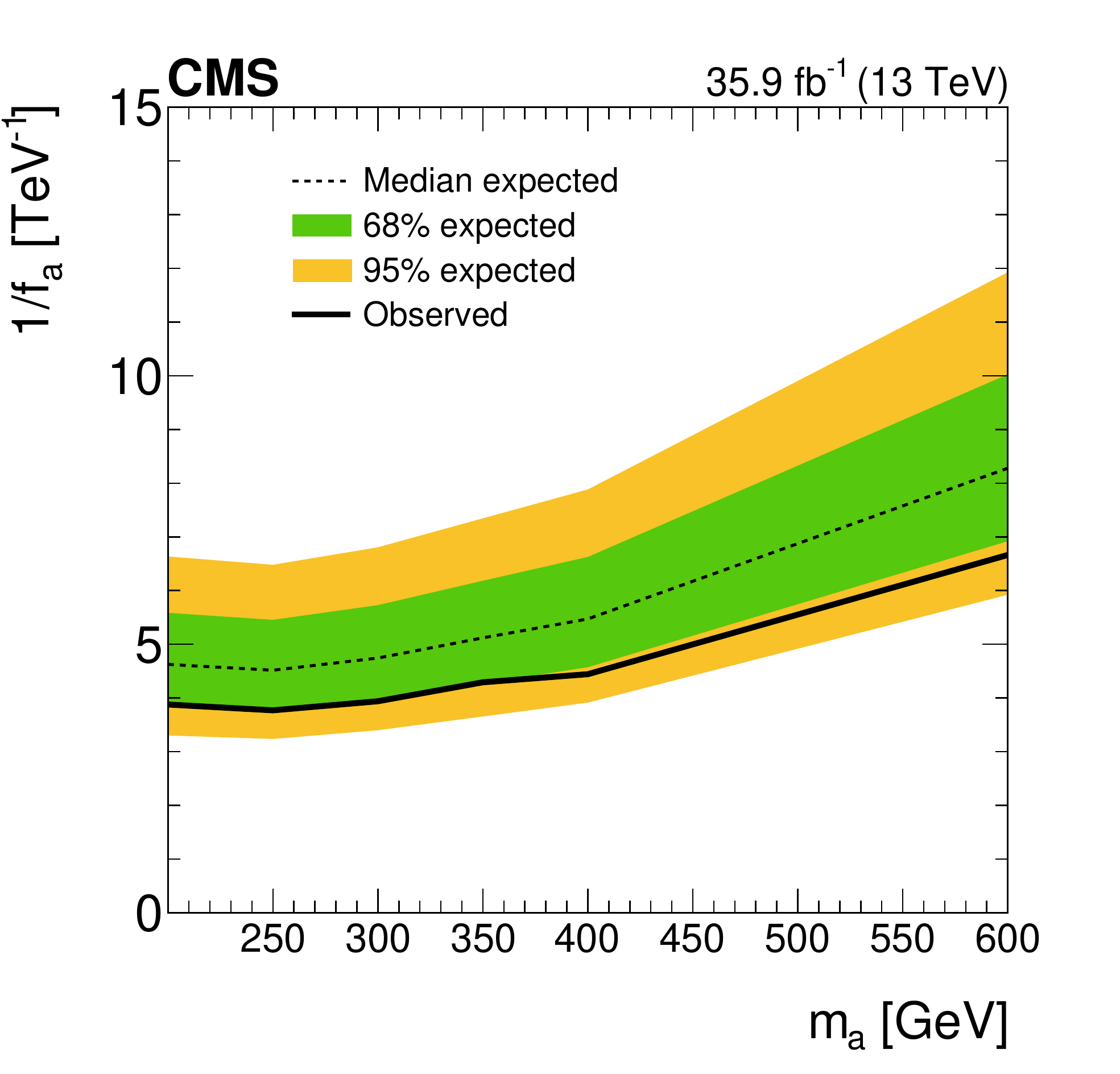}
    \caption{\label{fig:alplimit}
        (\cmsLeft)~Expected and observed 95\% \CL upper limits on the product of the cross section
        and branching fraction
        $\sigma(\Pp\Pp\to\PW {\Palp}) \mathcal{B}({\Palp}\to\WW)$
        as a function of ALP mass. The red line corresponds to the theoretical prediction for $1/\falp=5\TeV^{-1}$.
        (\cmsRight)~Expected and observed 95\% \CL upper limits on the photophobic ALP model parameter $1/\falp$
        as a function of ALP mass.
    }
\end{figure}

There is no evidence for an excess of events (Table~\ref{tab:results}).
Limits on the production of the
$\PW {\Palp}$ final state and on the parameter $1/\falp$ are placed using the methods described in
Section~\ref{sec:crosssection} for the SM production of \WWW.
The limits are displayed as a function of $\malp$ in Fig.~\ref{fig:alplimit}~(\cmsLeft) for
$\sigma(\Pp\Pp\to\PW {\Palp}) \mathcal{B}({\Palp}\to\WW)$ and
in Fig.~\ref{fig:alplimit}~(\cmsRight) for $1/\falp$.

\section{Summary\label{sec:summary}}

A search for \WWWpm production using proton-proton collision data at a center-of-mass energy of 13\TeV was presented.
Events with either two same-sign leptons (electrons or muons) and two jets or with three leptons with total charge $\pm$1
were selected.
The data were collected with the CMS experiment and correspond to an integrated luminosity of \theLumi.
The dominant sources of standard model backgrounds include nonprompt leptons, three-lepton events such as
those from the process $\WZ\to3\ell\nu$,
as well as \SSWW{}+ jets and \ttW production. Predictions for these backgrounds were derived or
validated using data in dedicated control regions.
The observed (expected) significance for \WWWpm production is 0.60 (1.78)~standard deviations and the ratio of measured signal yield to that
expected from the standard model is $0.34^{+0.62}_{-0.34}$, which corresponds to a measured cross section of $0.17^{+0.32}_{-0.17}\unit{pb}$.

New physics processes that could lead to an excess of events were considered. Limits on anomalous quartic gauge couplings
are set, for example; $-1.2 < \fTa / \Lambda^4 < 1.2\TeV^{-4}$ at 95\% confidence level. Limits are also set on the production of
axionlike particles in association with a \PW boson: mass points between $\malp = 200$ and~480\GeV are excluded
for the parameter value $1/\falp = 5\TeV^{-1}$.

\begin{acknowledgments}

We thank Nathaniel Craig and Skyler Kasko for performing calculations of the cross section for the ALP model and producing generator-level samples used in this analysis.

We congratulate our colleagues in the CERN accelerator departments for the excellent performance of the LHC and thank the technical and administrative staffs at CERN and at other CMS institutes for their contributions to the success of the CMS effort. In addition, we gratefully acknowledge the computing centers and personnel of the Worldwide LHC Computing Grid for delivering so effectively the computing infrastructure essential to our analyses. Finally, we acknowledge the enduring support for the construction and operation of the LHC and the CMS detector provided by the following funding agencies: BMWFW and FWF (Austria); FNRS and FWO (Belgium); CNPq, CAPES, FAPERJ, and FAPESP (Brazil); MES (Bulgaria); CERN; CAS, MoST, and NSFC (China); COLCIENCIAS (Colombia); MSES and CSF (Croatia); RPF (Cyprus); SENESCYT (Ecuador); MoER, ERC IUT, and ERDF (Estonia); Academy of Finland, MEC, and HIP (Finland); CEA and CNRS/IN2P3 (France); BMBF, DFG, and HGF (Germany); GSRT (Greece); OTKA and NIH (Hungary); DAE and DST (India); IPM (Iran); SFI (Ireland); INFN (Italy); MSIP and NRF (Republic of Korea); LAS (Lithuania); MOE and UM (Malaysia); BUAP, CINVESTAV, CONACYT, LNS, SEP, and UASLP-FAI (Mexico); MBIE (New Zealand); PAEC (Pakistan); MSHE and NSC (Poland); FCT (Portugal); JINR (Dubna); MON, RosAtom, RAS, RFBR and RAEP (Russia); MESTD (Serbia); SEIDI, CPAN, PCTI and FEDER (Spain); Swiss Funding Agencies (Switzerland); MST (Taipei); ThEPCenter, IPST, STAR, and NSTDA (Thailand); TUBITAK and TAEK (Turkey); NASU and SFFR (Ukraine); STFC (United Kingdom); DOE and NSF (USA).

\hyphenation{Rachada-pisek} Individuals have received support from the Marie-Curie program and the European Research Council and Horizon 2020 Grant, contract No. 675440 (European Union); the Leventis Foundation; the A. P. Sloan Foundation; the Alexander von Humboldt Foundation; the Belgian Federal Science Policy Office; the Fonds pour la Formation \`a la Recherche dans l'Industrie et dans l'Agriculture (FRIA-Belgium); the Agentschap voor Innovatie door Wetenschap en Technologie (IWT-Belgium); the Ministry of Education, Youth and Sports (MEYS) of the Czech Republic; the Council of Science and Industrial Research, India; the HOMING PLUS program of the Foundation for Polish Science, cofinanced from European Union, Regional Development Fund, the Mobility Plus program of the Ministry of Science and Higher Education, the National Science Center (Poland), contracts Harmonia 2014/14/M/ST2/00428, Opus 2014/13/B/ST2/02543, 2014/15/B/ST2/03998, and 2015/19/B/ST2/02861, Sonata-bis 2012/07/E/ST2/01406; the National Priorities Research Program by Qatar National Research Fund; the Programa Severo Ochoa del Principado de Asturias; the Thalis and Aristeia programs cofinanced by EU-ESF and the Greek NSRF; the Rachadapisek Sompot Fund for Postdoctoral Fellowship, Chulalongkorn University and the Chulalongkorn Academic into Its 2nd Century Project Advancement Project (Thailand); the Welch Foundation, contract C-1845; and the Weston Havens Foundation (USA).

\end{acknowledgments}

\bibliography{auto_generated}
\cleardoublepage \appendix\section{The CMS Collaboration \label{app:collab}}\begin{sloppypar}\hyphenpenalty=5000\widowpenalty=500\clubpenalty=5000\input{SMP-17-013-authorlist.tex}\end{sloppypar}
\end{document}

%% file: SMP-17-013-authorlist.tex
\vskip\cmsinstskip
\textbf{Yerevan Physics Institute, Yerevan, Armenia}\\*[0pt]
A.M.~Sirunyan$^{\textrm{\dag}}$, A.~Tumasyan
\vskip\cmsinstskip
\textbf{Institut für Hochenergiephysik, Wien, Austria}\\*[0pt]
W.~Adam, F.~Ambrogi, T.~Bergauer, J.~Brandstetter, M.~Dragicevic, J.~Erö, A.~Escalante~Del~Valle, M.~Flechl, R.~Frühwirth\cmsAuthorMark{1}, M.~Jeitler\cmsAuthorMark{1}, N.~Krammer, I.~Krätschmer, D.~Liko, T.~Madlener, I.~Mikulec, N.~Rad, J.~Schieck\cmsAuthorMark{1}, R.~Schöfbeck, M.~Spanring, D.~Spitzbart, W.~Waltenberger, C.-E.~Wulz\cmsAuthorMark{1}, M.~Zarucki
% ...
\vskip\cmsinstskip
\dag: Deceased\\
1:  Also at Vienna University of Technology, Vienna, Austria\\
% ...
